\documentclass[aps,prl, twocolumn,showpacs,reprint,superscriptaddress]{revtex4-1}

\usepackage{graphicx}
\usepackage{hyperref}
\usepackage{amsmath}
\usepackage{amssymb}

\begin{document}

\newcommand{\ket}[1]{$\left|#1\right\rangle$}

\title{Large Bragg Reflection from One-Dimensional Chains of Trapped Atoms\\Near a Nanoscale Waveguide}

\author{Neil V. Corzo}
\affiliation{Laboratoire Kastler Brossel, UPMC-Sorbonne Universit\'es, CNRS, ENS-PSL Research University, Coll\`ege de France, 4 Place
Jussieu, 75005 Paris, France}
\author{Baptiste Gouraud}
\affiliation{Laboratoire Kastler Brossel, UPMC-Sorbonne Universit\'es, CNRS, ENS-PSL Research University, Coll\`ege de France, 4 Place
Jussieu, 75005 Paris, France}
\author{Aveek Chandra}
\affiliation{Laboratoire Kastler Brossel, UPMC-Sorbonne Universit\'es, CNRS, ENS-PSL Research University, Coll\`ege de France, 4 Place
Jussieu, 75005 Paris, France}
\author{Akihisa Goban}
\thanks{Present address: JILA, University of Colorado, 440 UCB, Boulder, Colorado 80309, USA.}
\affiliation{Norman Bridge Laboratory of Physics 12-33, California Institute of Technology, Pasadena, CA 91125, USA}
\author{Alexandra S. Sheremet}
\affiliation{Information Technologies, Mechanics and Optics University, 199034 St. Petersburg, Russia}
\affiliation{Russian Quantum Center, 143025 Skolkovo, Moscow Region, Russia}
\author{Dmitriy V. Kupriyanov}
\affiliation{Department of Theoretical Physics, St. Petersburg State Polytechnic University, 195251 St. Petersburg, Russia}
\author{Julien Laurat}
\email{julien.laurat@upmc.fr}
\affiliation{Laboratoire Kastler Brossel, UPMC-Sorbonne Universit\'es, CNRS, ENS-PSL Research University, Coll\`ege de France, 4 Place
Jussieu, 75005 Paris, France}

\date{\today}

\begin{abstract}
We report experimental observations of large Bragg reflection from arrays of cold atoms trapped near a one-dimensional nanoscale waveguide. By using an optical lattice in the evanescent field surrounding a nanofiber with a period nearly commensurate with the resonant wavelength, we observe a reflectance of up to 75~\% for the guided mode. Each atom behaves as a partially-reflecting mirror and an ordered chain of about 2000 atoms is sufficient to realize an efficient Bragg mirror. Measurements of the reflection spectra as a function of the lattice period and the probe polarization are reported. The latter shows the effect of the chiral character of nanoscale waveguides on this reflection. The ability to control photon transport in 1D waveguides coupled to spin systems would enable novel quantum network capabilities and the study of many-body effects emerging from long-range interactions.
\end{abstract}

\maketitle

In recent years, the coupling of one-dimensional bosonic waveguides and atoms, either real or artificial, has raised a large interest \cite{Chang2007,vanLoo2013,Lodahl2015}. Beyond the remarkable ability to couple a single emitter to a guided mode \cite{Lodahl2015}, the 1D reservoir would also enable the exploration and eventual engineering of photon-mediated long-range interactions between multiple qubits, a challenging prospect in free-space geometries. This emerging field of waveguide quantum electrodynamics promises unique applications to quantum networks, quantum nonlinear optics and quantum simulation \cite{Kimble,ChangReview,Gonzalez2015}.

In this context, progress has been reported on various fronts. In the microwave regime, the coupling of superconducting qubits to a one-dimensional transmission line provides a versatile platform to study such photon-mediated interactions \cite{vanLoo2013}. At optical frequencies, recent experimental advances include the development of 1D nanoscale dielectric waveguides coupled to cold atoms trapped in the vicinity \cite{nanofiber1, nanofiber2, phc}. In these experiments, tight transverse confinement of the electric field achieves an effective mode area comparable to the atomic cross section and thereby a strong atom-photon interaction in a single-pass configuration \cite{Domokos2002}.

Coupling of atom arrays to 1D waveguides could lead to a variety of remarkable cooperative phenomena. This coupling can strongly modify the photon transport properties \cite{Shen2005,Dzsotjan2010,LeKien2014}, resulting for instance in sub- and superradiant decays as recently observed for two coupled atoms \cite{Goban2015}. It can also lead to photonic bandgaps and provide atomic Bragg mirrors, with envisioned applications to integrated cavity-QED \cite{Yue2011,Chang2012, Guimond2016}. This setting is as well at the basis of a recently proposed deterministic state engineering protocol \cite{Kimble2015} and constitutes the building block of chiral spin networks in which the emission into the left- and right-propagating modes is asymmetric \cite{Pichler2015}. Moreover, strong optomechanical couplings resulting from photon-mediated forces can give rise to rich spatial atomic configurations, including self-organization \cite{Asboth2008,Chang2013}.

Optical nanofibers offer a promising platform for exploring these effects. Their subwavelength diameter results in a large evanescent field that can be used for both trapping and interacting with atoms \cite{Sile}. Seminal works achieved the trapping of cold atoms near an optical nanofiber \cite{nanofiber1, nanofiber2} and recently enabled the demonstrations of all-fibered optical memories \cite{Gouraud2015,Sayrin2015}. These works were realized with disordered atoms or incommensurate arrays, and relied on the optical depth of the medium. An important capability that has not been demonstrated heretofore is the realization of cooperative effects emerging from the spatial order of the atoms, when the lattice is commensurate or close-to-commensurate with the resonant wavelength. 

In this Letter, we report a large Bragg reflection from atom arrays near a one-dimensional waveguide in a situation where each trapping site is occupied by at most one atom. Due to the tight transverse confinement of the guided light, a few thousand atoms trapped in the evanescent field of a nanofiber are sufficient to strongly reflect the incoming light. By using a near-resonant dipole trap, the maximum Bragg reflection is obtained for a slightly detuned probe. We provide detailed characterizations of the reflection spectra and finally show the effect of the waveguide chirality arising from the complex polarization pattern of tightly focused light. 

The long-range order of trapped atoms can indeed dramatically change the scattering properties. In incommensurate arrays, the interference between forward- and backward-scattered light leads to absorption and a vanishing reflection. In contrast, in commensurate arrays, this interference can result in strong reflection close to resonance \cite{Deutsch1995}. This effect, well-known as Bragg reflection, has been largely studied in crystals as well as in multilayer dielectric structures. It has also been observed with ordered cold atoms in free-space, either with three-dimensional \cite{Weidemuller1995,Birkl1995} or one-dimensional optical lattices \cite{Slama2005,Schilke2011}. A reflectance as high as 80\% was demonstrated in \cite{Schilke2011}. This observation required around $10^7$ atoms distributed over 7700 layers to reach the regime of multiple reflections. In contrast, in this work we demonstrate that 2000 atoms are sufficient to achieve large reflectance in a waveguide-mediated scenario.

\begin{figure}[t!]
\includegraphics[width=0.97\columnwidth]{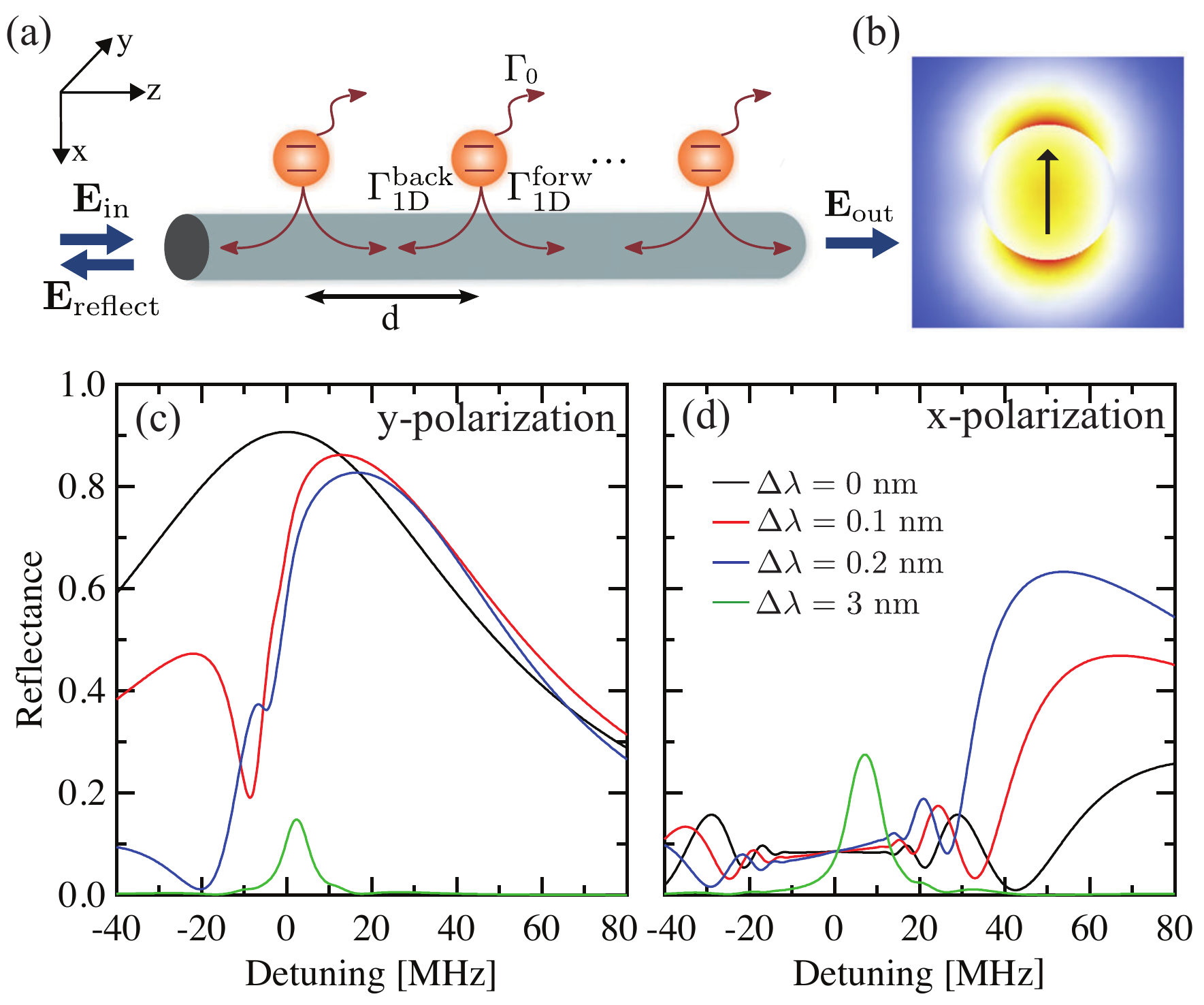}
\caption{(color online). Bragg reflection from atoms coupled to a one-dimensional waveguide. (a) $N$ atoms are trapped near of a waveguide and exhibit radiative decay rates $\Gamma_{\textrm{1D}}^{\textrm{forw,back}}$ into the right and left-propagating modes, and $\Gamma'\simeq\Gamma_0$ into all the other modes. (b) Electric field distribution in the transverse plane of a nanofiber for a guided probe with a quasilinear polarization (indicated by the arrow). (c) Theoretical reflection spectra for a probe quasilinearly polarized along the y-direction (symmetric decay rates) and (d) along the x-direction (asymmetric decay rates). The spectra are given for different distances between the atoms, with values close to the commensurate case. $\Delta\lambda$ stands for the trap detuning to resonance, with $d=\lambda_0/2+\Delta\lambda/2$. ($N=2000$, $\Gamma_{\textrm{1D}}/\Gamma_0=0.01$, $\Gamma_{\textrm{1D}}^{\textrm{forw}}=2.8\Gamma_{\textrm{1D}}$, $\Gamma_{\textrm{1D}}^{\textrm{forw}}/\Gamma_{\textrm{1D}}^{\textrm{back}}=12$). Theoretical values of the couplings are taken from \cite{LeKien_chiral}.}
\label{fig1}
\end{figure}

To introduce the scattering properties and associated photon transport, we first consider a typical configuration as depicted in Fig. \ref{fig1}(a). $N$ atoms are trapped in the vicinity of a waveguide, with a lattice constant $d$ close to $\lambda_0/2$, where $\lambda_0$ corresponds to wavelength of the atomic transition. Due to the complex polarization structure of tightly focused light \cite{Balykin2004}, which includes a significant longitudinal component, the scattering in the guided mode can be asymmetric \cite{LeKien_chiral}. Each atom exhibits a radiative decay rate $\Gamma_{\textrm{1D}}^{\textrm{forw}}$ and $\Gamma_{\textrm{1D}}^{\textrm{back}}$ into the right- and left-propagating mode respectively, and $\Gamma'\simeq\Gamma_0$ into all the other modes. $\Gamma_0$ is the radiative decay rate in free space. For a guided probe field quasilinearly polarized along the y-direction, the two decay rates are equal, $\Gamma_{\textrm{1D}}^{\textrm{back}}=\Gamma_{\textrm{1D}}^{\textrm{forw}}=\Gamma_{\textrm{1D}}/2$. For an orientation along the x-direction, i.e. pointing towards the atoms, the couplings to the waveguide become strongly asymmetric. It has also been shown that when the ground state Zeeman levels are equally populated, the two x- and y-polarization are not coupled to each other by the linear coherent scattering \cite{LeKien_bragg}. For nanofiber-trapped atoms, typically around 200 nm from the surface, the ratio $P=\Gamma_{\textrm{1D}}/\Gamma_0$ amounts to a few $10^{-2}$. In the case of asymmetric coupling, the forward decay rate is increased by sixfold while the backward decay rate is suppressed by about one order of magnitude. 

\begin{figure}[b!]
\includegraphics[width=0.97\columnwidth]{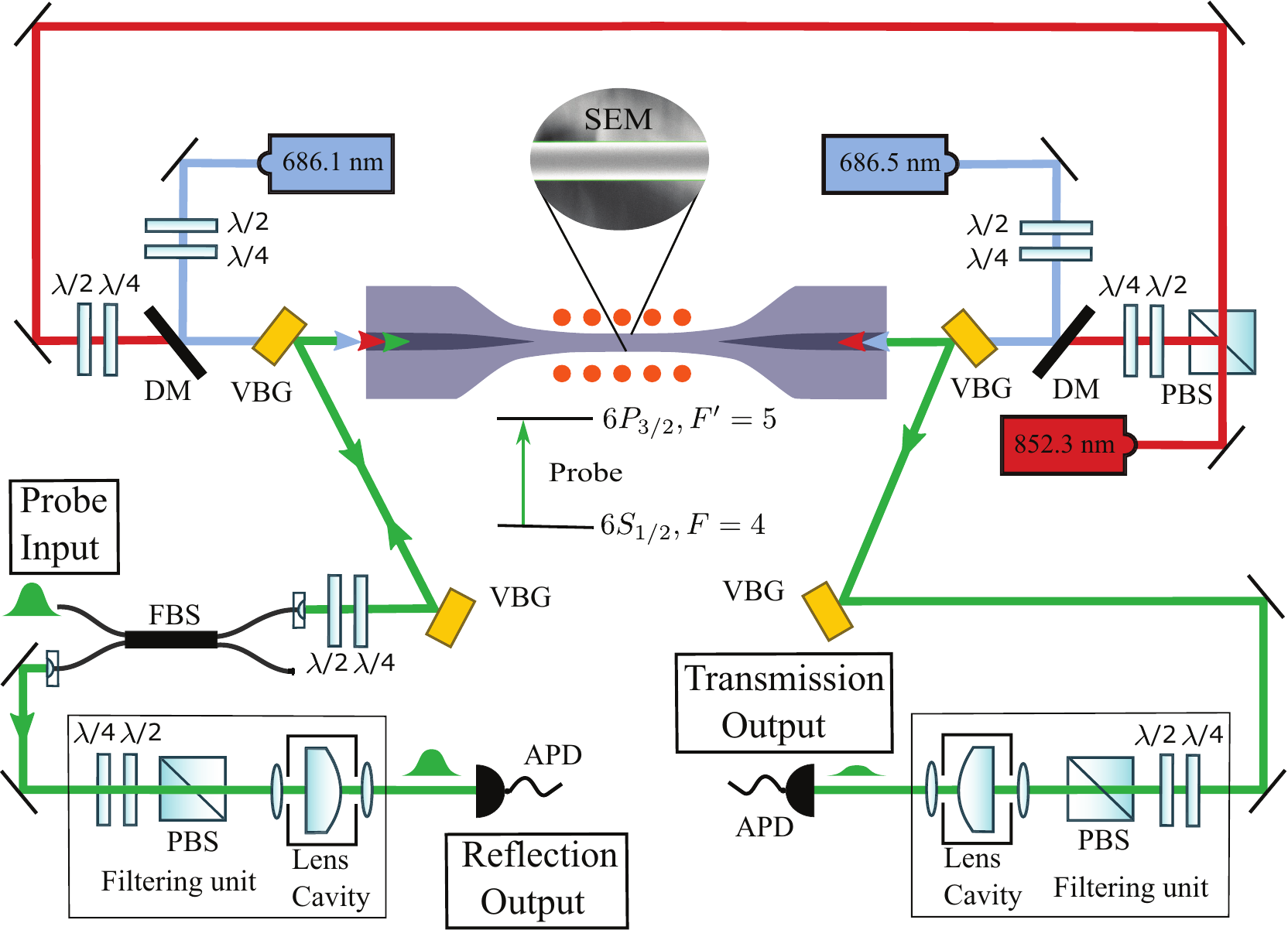}
\caption{(color online). Experimental setup. Arrays of cesium atoms are trapped in the evanescent field of a nanofiber. The lattice is realized by a pair of close-to-resonance red-detuned  counterpropagating beams. An additional pair of blue-detuned beams with slightly different wavelengths gives a repulsive contribution. The atoms are prepared in the $|g\rangle=\{\rm6S_{1/2}, F=4\}$ ground state and the probe addresses the transition $|g\rangle \rightarrow |e\rangle=\{\rm6P_{3/2}, F'=5\}$. Around 2000 atoms in total are trapped along two parallel chains. Reflection and transmission spectra are measured with avalanche photodiodes after polarization and frequency filterings. DM stands for dichroic mirror, FBS for fiber beamsplitter and VBG for volume Bragg grating.}
\label{fig2}
\end{figure}

The single-atom amplitude reflection $r$ can be calculated from the couplings to the waveguide and the probe detuning $\delta$ to resonance. In the symmetric decay case for instance, the reflection is given by $r\simeq-\Gamma_{\textrm{1D}}/(\Gamma_{0}-2i\delta)$  \cite{Chang2007}. At resonance, one atom thus reflects $P^2\simeq 10^{-4}$ in intensity. The transfer matrix formalism \cite{Deutsch1995} enables then to calculate the single-photon propagation through the ensemble (see appendix). 

Figure \ref{fig1}(c) and \ref{fig1}(d) provide theoretical reflection spectra for different small detunings $\Delta\lambda$ of the trap wavelength to atomic resonance and for the two orthogonal polarizations. For atoms separated exactly by $\lambda_0/2$, the reflection spectrum is a broadened Lorentzian in the symmetric coupling case while the reflectance is strongly suppressed in the chiral one. Indeed, the amount of chirality and number of atoms result in a finite bandwidth around resonance where reflection is suppressed, as detailed in \cite{LeKien_bragg}. For close-to-commensurate traps, as studied here, the Bragg condition is fulfilled out of resonance. This leads to a maximum reflectance shifted to the blue \cite{Schilke2011} but also results in an increased reflectance for the chiral case. Large reflectance values can then be obtained for both polarizations as the single-atom reflection coefficients are similar in our configuration.

Our experimental setup is illustrated in Fig. \ref{fig2}. Arrays of trapped atoms are prepared in the evanescent field of a 400-nm diameter nanofiber suspended inside an ultrahigh vacuum chamber. The nanofiber is produced from a single-mode fiber (OZ Optics SMF-780-5/125) by the standard heating-pulling technique. The polarization of the different guided beams can be aligned by measuring the polarization properties of the light scattered from imperfections at the fiber surface \cite{Vetsch12,nanofiber2}.

Our two-color dipole trap is based on the combination of laser beams all guided by the nanofiber. By using the appropriate combination of attractive red-detuned light and repulsive blue-detuned light, the atoms are trapped in the potential minima located at a sub-wavelength distance from the waveguide \cite{Balykin, nanofiber1, nanofiber2}. The longitudinal periodic structure is given by the standing wave formed by the two counter-propagating red-detuned beams, with $\Delta\lambda$ the detuning from resonance. We perform our experiment at two different values of $\Delta\lambda$, 0.12 nm and 0.2 nm, with a total power equal to $2\times 1\text{ }\mu$W and $2\times 1.9\text{ }\mu$W respectively. A pair of counter-propagating beams ($2\times 4$ mW) at 686.1 nm and 686.5 nm are chosen as blue-detuned beams and contribute to the trap in a compensated manner \cite{nanofiber2}. The polarization of the beams is oriented along the transverse $x$-direction. The resulting intensity pattern leads to trapping sites aligned along two lines parallel to the fiber. Because of the loading dynamics, each site hosts one atom at most \cite{Schlosser02}. Simulations of the trapping potential (see appendix) provide for $\Delta\lambda= 0.12$ nm (0.2 nm) a trap depth at minimum equal to -0.15 mK (-0.1 mK) and an axial trap frequency $\nu_z/2\pi=258$ kHz (215 kHz). For $\Delta\lambda= 0.2$ nm, we measure a shift of the transition equal to 3 MHz relative to free-space and an inhomogeneous broadening limited to $\sigma=0.6\Gamma_0$.
 
 \begin{figure}[t!]
\includegraphics[width=0.95\columnwidth]{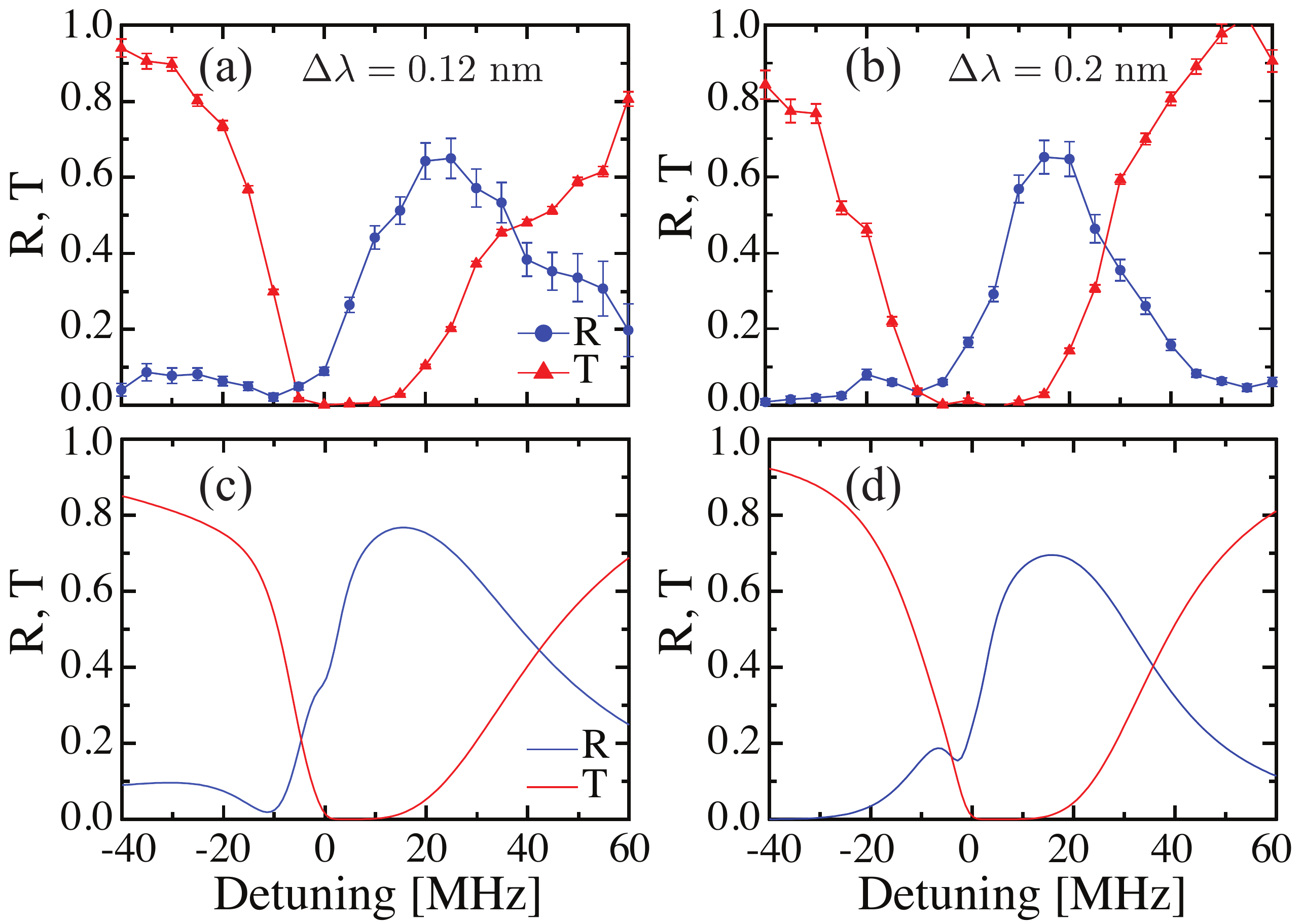}
\caption{(color online). Reflection and transmission spectra for a probe quasilinearly polarized along the y-direction. (a) and (b) Experimental results for $\Delta\lambda=$0.12 nm and $\Delta\lambda=$0.2 nm. (c) and (d) Simulated spectra for $N=2000$ atoms, $\Gamma_{\textrm{1D}}/\Gamma_0=0.007$ and a filling factor of the lattice sites $f=0.3$. The coupling value and filling factor have been adjusted to fit the measured spectra reported in (a).}
\label{fig3}
\end{figure}

To load the dipole trap, which is continuously on, we overlap a cigar-shaped magneto-optical trap (MOT) along the fiber axis. The atomic cloud is then further cooled and the lattice is loaded during an optical molasses phase: the MOT coils are switched off, and during a 10 ms interval, the MOT cooling-beam detuning is increased from $-2\Gamma_0$ to $-16\Gamma_0$, while the powers of both the cooling and repumping beams are decreased to zero. The atoms are then optically pumped into the $|g\rangle=\{\rm6S_{1/2}, F=4\}$ hyperfine ground state for $\rm 400\text{ }\mu s$. By a saturation measurement, we estimate the number of trapped atoms  to be $N=2000\pm200$. Residual magnetic fields are canceled with bias coils and characterized via Zeeman sublevel microwave spectroscopy. The fields are compensated down to the 20 mG level. 

To measure the reflection and transmission spectra, a probe pulse first passes through a fibered beamsplitter that we use to separate the reflected pulse from the input probe. The probe is then combined with the pair of trapping beams, via a Volume Bragg Grating (VBG), and sent into the nanofiber. The weak probe pulse arriving on the atoms has a mean photon number of 2$\pm$0.05. The transmitted pulse is filtered by another VBG and an additional filtering system, and finally detected by an avalanche photodiode. The filtering system combines a polarization beamsplitter and a commercial lens-based cavity \cite{lens}. The cavity transmission is around 75\%, with a bandwidth of 80 MHz and a rejection around 40 dB for the trapping beams. These cascaded stages are required to filter out efficiently the dipole beams and reach the single-photon level. The reflected pulse is directed to a similar filtering and detection stage. 

The reflectance values are obtained by comparing the reflected pulse when atoms are trapped to the transmitted pulse without atoms, and by correcting for the different losses in the system. These losses, which are detuning-dependent due to the filtering cavities, are calibrated before and after each measurement. The transmittance values are obtained by comparing the transmitted pulse with and without trapped atoms. Error bars in the data include the calibration uncertainties.

\begin{figure}[t!]
\includegraphics[width=0.96\columnwidth]{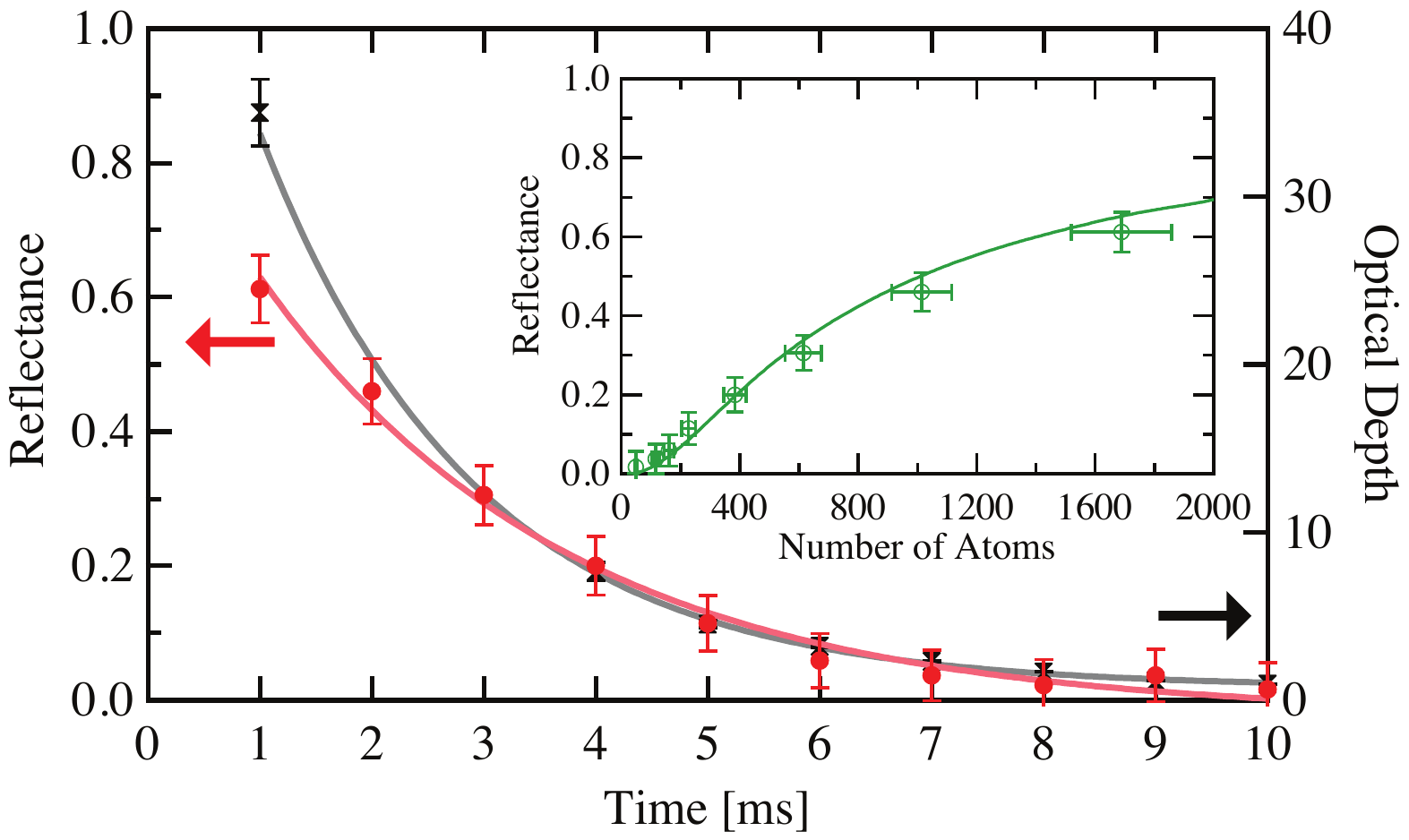}
\caption{(color online). Decay of the reflectance with trapping time and atom number. The largest measured reflectance (red) and the optical depth of the medium at resonance (black) are given as a function of the trapping time. The solid lines are exponential decay fits. The inset provides the reflectance as a function of the remaining number of trapped atoms. The green line is given by a simple model taking into account an initial filling factor $f=0.3$ and subsequent random loss ($\Delta\lambda=$0.2 nm, $\Gamma_{\textrm{1D}}/\Gamma_0=0.007$).}
\label{fig4}
\vspace{-0.5cm}
\end{figure}

We first characterize the reflection and transmission spectra for a probe quasilinearly polarized along the y-direction. In this configuration, the radiative decay rates are expected to be symmetric. The measurements are given in Fig. \ref{fig3} for the two different detunings of the dipole trap. For $\Delta\lambda=$0.12 nm, as shown in Fig. \ref{fig3}(a), we observe a broad reflection spectrum, with a maximum reflectance of (0.65$~\pm~$0.05) for a detuning of 25 MHz.  For $\Delta\lambda=$0.2 nm, the reflection peak is narrower and the maximum reflectance (0.65$~\pm$0.04) is shifted to a lower frequency (Fig. \ref{fig3}(b)). 

The spectra resulting from the two trapping distances are well explained by our simple theoretical model, as presented in Fig. \ref{fig3}(c) and Fig. \ref{fig3}(d). In this model, we take into account the disorder in the distribution of the atoms along the two parallel arrays due to a limited filling factor $f$ of the individual trapping sites (see appendix). The coupling to the waveguide and the filling factor have been adjusted to provide the best fit to the spectra of Fig. \ref{fig3}(a). A value $f=0.3$ was obtained. Larger values lead to a pronounced dip on the reflection spectrum while smaller values result in a strongly reduced reflectance and a narrowing of the spectra width. The simulations take also into account the shift induced by the dipole trapping. Inhomogeneous broadening is not included as the limited broadening $\sigma$ leads to a negligible modification of the spectra (see appendix). 

Apart from a finite filling factor, a second contribution to disorder can arise from the thermal distribution of the atoms in the potential wells. The good agreement between the achieved reflectances and the predicted ones supports a tight axial localization, as expected with our microscopic traps. With a temperature $T=20\mu K$ estimated by a time-of-flight measurement and the predicted axial trap frequency of $\nu_z/2\pi=258$ kHz, the spread is given in the harmonic approximation by $\sigma_z=\sqrt{(k_b T)/(m_{\textrm{Cs}}\nu_z^2)}\sim 22$ nm, i.e. $\sim\lambda/40$. This value leads to a Debye-Waller factor $f_{DW}=e^{-4k^2\sigma_z^2}$ close to 0.9 \cite{fdw}. As shown in the appendix, this disorder has a very limited effect in our configuration where Bragg reflection is observed out of resonance \cite{Slama2006}. 

\begin{figure}[t!]
\includegraphics[width=0.93\columnwidth]{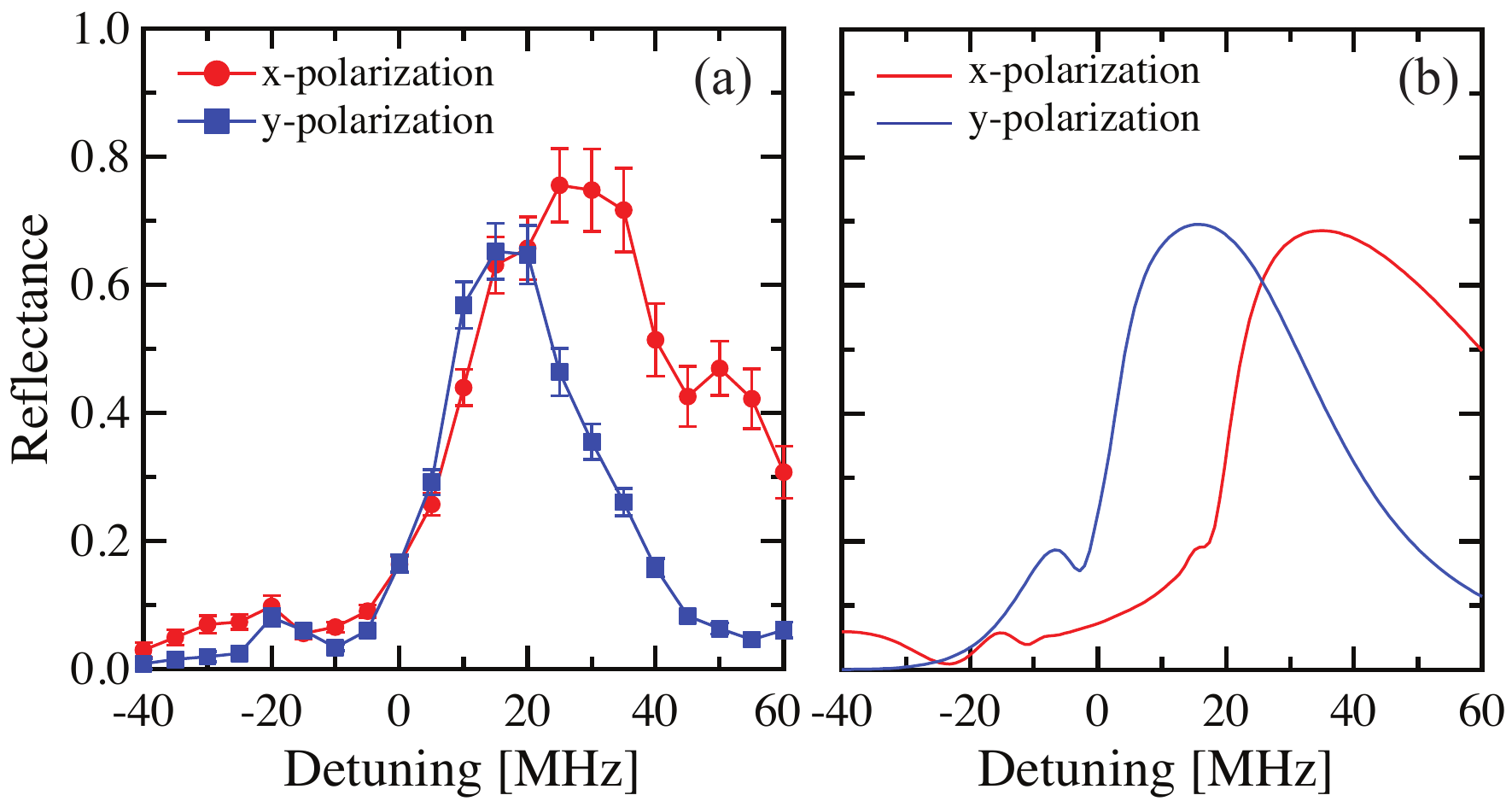}
\caption{(color online). Effect of the chiral character of the waveguide on Bragg reflection. (a) Measured reflection spectra for x- and y-quasilinear polarization, with $\Delta\lambda=$0.2 nm. (b) Theoretical simulations ($N=2000$, $\Gamma_{\textrm{1D}}/\Gamma_0=0.007$, $\Gamma_{\textrm{1D}}^{\textrm{forw}}=2.8\Gamma_{\textrm{1D}}$, $\Gamma_{\textrm{1D}}^{\textrm{forw}}/\Gamma_{\textrm{1D}}^{\textrm{back}}=12$, $f=0.3$).}
\label{fig5}
\end{figure}

The measurements described above were performed 1 ms after loading the trap. We now investigate the reflection for longer trapping time and compare its decay with the trap lifetime. Figure \ref{fig4} shows the maximal reflectance and optical depth of the medium as a function of the hold time. Using exponential fits, we obtain a decay time of 2.7 ms and 1.9 ms respectively. From the measured optical depth, we then estimate the number of remaining trapped atoms at each time. The inset in Fig.~\ref{fig4} finally provides the maximal reflectance as a function of this inferred atom number. The green line takes into account a random loss of the atoms from the initial arrays with a total atom number of 2000 and a filling factor $f=0.3$ (see appendix). This model agrees well with the data. 

Finally, we investigate the effect of the chiral character of the waveguide. The previous measurements were realized with the y-polarization, which leads to symmetric decay rates for the backward and forward scattering. We now consider the asymmetric case. This situation can be obtained with a probe quasilinearly polarized along the x-direction. The measured reflection spectra for both probe polarizations are compared in Fig. \ref{fig5}(a). The trap detuning is fixed at $\Delta\lambda=$0.2 nm. As it can be seen, the spectrum is significantly shifted and broadened in the asymmetric case. These features are compelling signatures of the chiral character of the waveguide on the reflection, as confirmed by the associated simulations in Fig.~\ref{fig5}(b). The maximal observed reflectance of (0.75$~\pm~$0.06) is obtained in this asymmetric case, at a probe detuning of 25 MHz.

In conclusion, we have realized an efficient Bragg atomic mirror based on a nanoscale one-dimensional waveguide coupled to about 2000 atoms. The effect of the chiral character of the waveguide on the reflection features has also been observed. Beyond their fundamental significance, these observations demonstrate key ingredients for the exploration of a variety of emerging and potentially rich protocols based on 1D reservoirs coupled to atoms. To enable an exact commensurate array of trapped atoms, and therefore an enhanced reflection closer to resonance, a bichromatic optical superlattice, whose trap periodicity is given by the tunable beat frequency of the trapping lasers \cite{Gorlitz2001,Camposeo2003}, can be developed. Such a trapping scheme, which has been demonstrated in free space, would need to be adapted to the evanescent field configuration around a nanofiber. This superlattice could also include a double primitive cell enabling a richer photonics spectrum \cite{Morigi2009}. An additional classical driving field in a three-level atomic configuration would finally provide a dynamical control of the transport properties \cite{Witthaut2010}.\\

\textit{Note -- During the preparation of this manuscript, a related experiment by H.L. S$\o$rensen and coworkers was presented in \cite{Appel2016}.}\\

\begin{acknowledgments}
This work is supported by the European Research Council (Starting Grant HybridNet), the Emergence program from Ville de Paris and the IFRAF DIM NanoK from R\'egion Ile-de-France. The authors acknowledge interesting discussions within the CAPES-COFECUB project Ph 740-12 and with H.J. Kimble when J.L. was in a visiting professor position at Caltech in 2015. The authors also thank A. Nicolas, D. Maxein and O. Morin for their contributions in the early stage of the experiment, and A.C. McClung for constructive comments on the manuscript. A.S. and D.V.K. acknowledge RFBR (Grant No. 15-02-01060). N.V.C. is supported by the EU (Marie Curie Fellowship). J.L. is a member of the Institut Universitaire de France.
\end{acknowledgments}

\clearpage

\appendix

\section{S1. Spectra simulation}

\begin{figure}[b!]
\includegraphics[width=0.97\columnwidth]{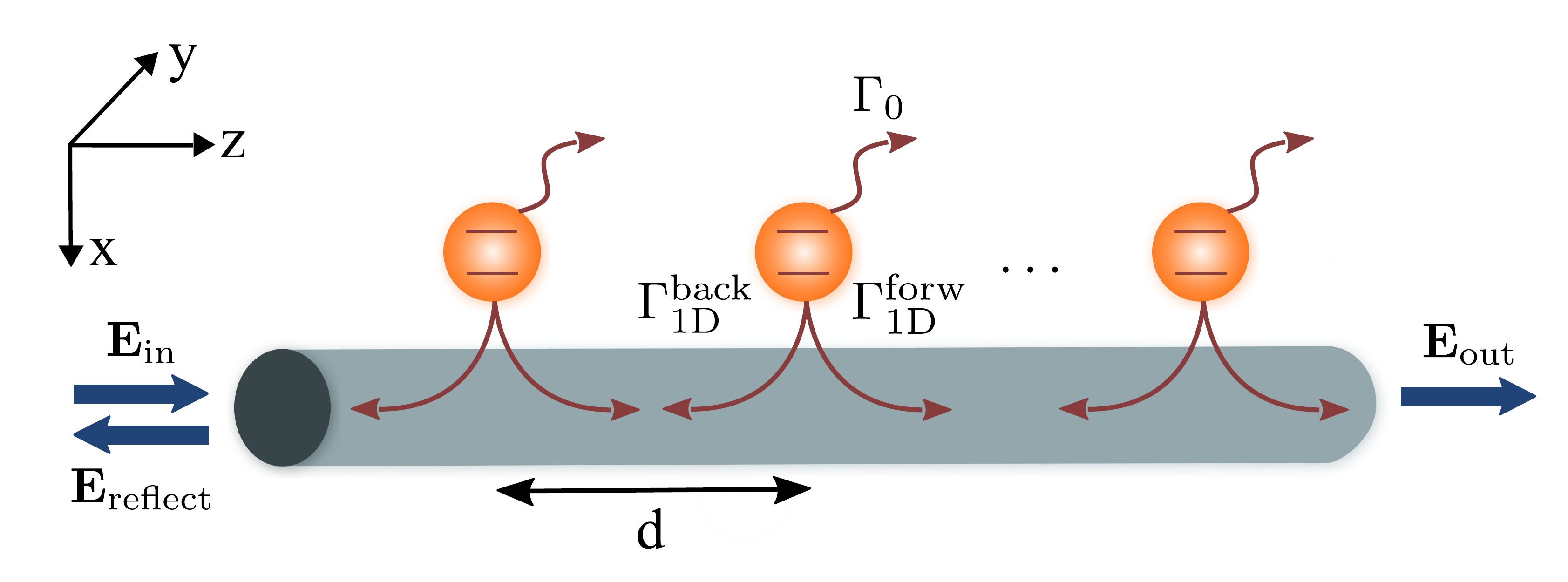}
\caption{ $N$ atoms are trapped in the vicinity of a nanoscale waveguide with radiative decay rates $\Gamma_{\textrm{1D}}^{\textrm{forw,back}}$ into the right and left-propagating modes and $\Gamma'\simeq\Gamma_0$ into all the other modes.}
\label{fig1_SI}
\end{figure}

The light propagation through a one-dimensional chain, as illustrated in Fig. \ref{fig1_SI}, can be described by the transfer matrix formalism \cite{Deutscha}. For an atom at position $z$ one can introduce the matrix $M_a$ that relates components of the backward- and forward-traveling electric field on the right side $E_R^{(back)}$ and $E_R^{(forw)}$ to the ones on the left side $E_L^{(back)}$ and $E_L^{(forw)}$:
\begin{equation}
\begin{pmatrix}  E_L^{(back)}\\ E_L^{(forw)} \end{pmatrix} = M_a \cdot \begin{pmatrix}  E_R^{(back)}\\ E_R^{(forw)} \end{pmatrix}.
\label{M_a-matrix}
\end{equation}
The boundary conditions for the wave equation allows to obtain the following expression for the matrix (\ref{M_a-matrix}):
\begin{equation}
M_a = \frac{1}{t}\begin{pmatrix}  t^2 - r^2 & r \\ -r & 1 \end{pmatrix} ,
\label{M_a-matrix2}
\end{equation}
where $t$ and $r$ are single-atom transmission and reflection coefficients respectively.
The transfer matrix $M_p$ for propagation between neighboring atoms is simply given by
\begin{equation}
M_p = \begin{pmatrix}  e^{ikd} & 0 \\ 0 & e^{-ikd} \end{pmatrix} ,
\label{M_t-matrix}
\end{equation} 
where $d$ is the distance between the two atoms. The transfer matrix for the full ensemble is obtained as a product of matrices $M_a$ and $M_p$ as follows:
\begin{equation}
M = \begin{pmatrix}  M_{11} & M_{12} \\ M_{21} & M_{22} \end{pmatrix} = (M_a \cdot M_p)^N.
\label{M-matrix}
\end{equation}
The transmission and reflection coefficients of the atomic chain are finally given by:
\begin{eqnarray}
T &=& \left| \frac{1}{M_{22}}\right| ^2
\nonumber\\
R &=& \left| \frac{M_{12}}{M_{22}}\right| ^2.
\label{TR}
\end{eqnarray}

The single-atom transmission and reflection coefficients $t$ and $r$ depend on the radiative decay rates. Each atom has two channels of decay $\Gamma = \Gamma_{1D} + \Gamma'$ that can be associated with the guided mode $\Gamma_{1D}$ and the radiation modes $\Gamma'\sim \Gamma_0$ where $\Gamma_0$ is the radiative decay in free-space. The decay to the guided mode can occur in forward $\Gamma_{1D}^{forw}$ and backward $\Gamma_{1D}^{back}$ directions. Given the polarization of the guided light and positions of the atoms, $\Gamma_{1D}^{forw}$ and $\Gamma_{1D}^{back}$ can be equal or asymmetric \cite{LeKien_chirala}. 

In the case of quasilinear polarization along $y$, i.e. orthogonal to the atomic array, the decay rates in the forward and backward directions are equal $\Gamma_{1D}^{forw} = \Gamma_{1D}^{back} = \Gamma_{1D}/2$. In this symmetric case, the reflection and transmission coefficients can be expressed as \cite{Changa}:
\begin{eqnarray}
r &=& -\frac{\Gamma_{1D}}{\Gamma - 2i\delta}
\nonumber\\
t &=& 1+ r
\end{eqnarray}
where $\delta$ is the detuning from the atomic resonance. In accordance with (\ref{M_a-matrix2}) the explicit expression of the transfer matrix $M_a$ can be written as follows:
\begin{equation}
M_a = \frac{1}{\Gamma_0 - 2i\delta}\begin{pmatrix}  \frac{(\Gamma_0 - 2i\delta)^2 - \Gamma_{1D}^2}{\Gamma - 2i\delta} & -\Gamma_{1D} \\ \Gamma_{1D} & \Gamma - 2i\delta \end{pmatrix}.
\label{M_a_ortho}
\end{equation}

In the case of asymmetric scattering in forward and backward directions, a situation that occurs if the guided mode is quasilinearly polarized along the $x$-direction (pointing towards the atomic chain), the full decay rate can be written as $\Gamma = \Gamma' + \Gamma_{1D}^{forw} + \Gamma_{1D}^{back}$.
The explicit expressions for these coefficients can be found in accordance with \cite{Kiena} as follows:
\begin{eqnarray}
r &=& - \frac{2\sqrt{\Gamma_{1D}^{forw}\cdot \Gamma_{1D}^{back}}}{\Gamma - 2i\delta}
\nonumber\\
t &=& 1 - \frac{(\Gamma_{1D}^{forw} + \Gamma_{1D}^{back})}{\Gamma - 2i\delta}
\end{eqnarray}
These expressions finally provide the transfer matrix (\ref{M_a-matrix2}) for the asymmetric case as:
\begin{equation}
M_a = \frac{1}{\Gamma_0 - 2i\delta}\begin{pmatrix}  \frac{(\Gamma_0 - 2i\delta)^2 - 4\Gamma_{1D}^{forw}\cdot\Gamma_{1D}^{back}}{\Gamma - 2i\delta} & -2\sqrt{\Gamma_{1D}^{forw}\cdot\Gamma_{1D}^{back}} \\ 2\sqrt{\Gamma_{1D}^{forw}\cdot\Gamma_{1D}^{back}} & \Gamma - 2i\delta \end{pmatrix} ,
\label{M_a-matrix_par}
\end{equation}

\section{S2. Effect of inhomogeneous broadening}

The dipole trapping of the atoms can result in a resonance shift and an inhomogeneous broadening. This effect is limited in the reported experiment: the shift has been measured equal to 3 MHz and the broadening to $\sigma_{\delta}\sim$ 3 MHz. To study the effect of these parameters, we vary randomly the detuning for each atom of the chain, $\delta_i = \delta + \delta_{b_i}$ in accordance with the Gaussian distribution:
\begin{equation}
f(\delta_b) = \frac{1}{\sqrt{2\pi \sigma_{\delta}^2}}e^{-\frac{\delta_b^2}{2\sigma_{\delta}^2}}
\end{equation}
where $\sigma_{\delta}$ is the standard deviation. Each atom has an individual transfer matrix $M_{a_{i}}$ and the total transfer matrix for the chain can be read as:
\begin{equation}
M = (M_{a_1}\cdot M_p)\cdot (M_{a_2}\cdot M_p)\cdot ... \cdot (M_{a_N}\cdot M_p)
\end{equation}

\begin{figure}[t!]
\includegraphics[width=0.95\columnwidth]{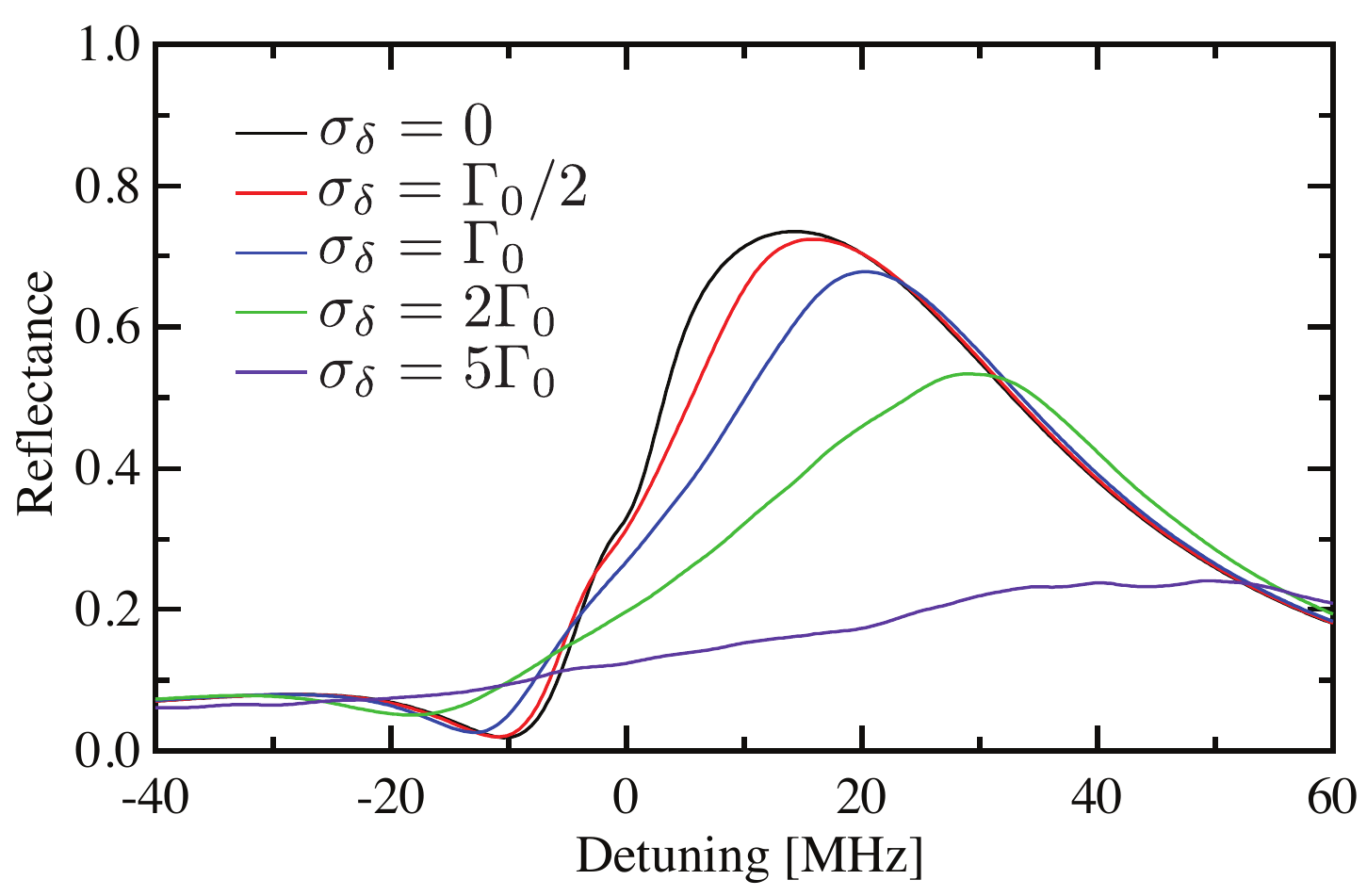}
\caption{Simulated reflection spectra for a probe quasilinearly polarized along the y-direction (symmetric decay rates) and different inhomogeneous broadening $\sigma_{\delta}$. The trap detuning is $\Delta\lambda=0.2$ nm, the number of atoms $N=2000$ and the decay rate $\Gamma_{\textrm{1D}}/\Gamma_0=0.007$. The spectra are averaged over 15 realizations.}
\label{fig2_SI}
\end{figure}

In the main manuscript, simulations include the shift but not the broadening. The measured broadening has indeed small effect, below what can be measured given the precision of our measurements. Figure \ref{fig2_SI} provides the simulations of the reflexion spectra for different broadenings $\sigma_{\delta}$. 

\begin{figure}[b!]
\includegraphics[width=0.95\columnwidth]{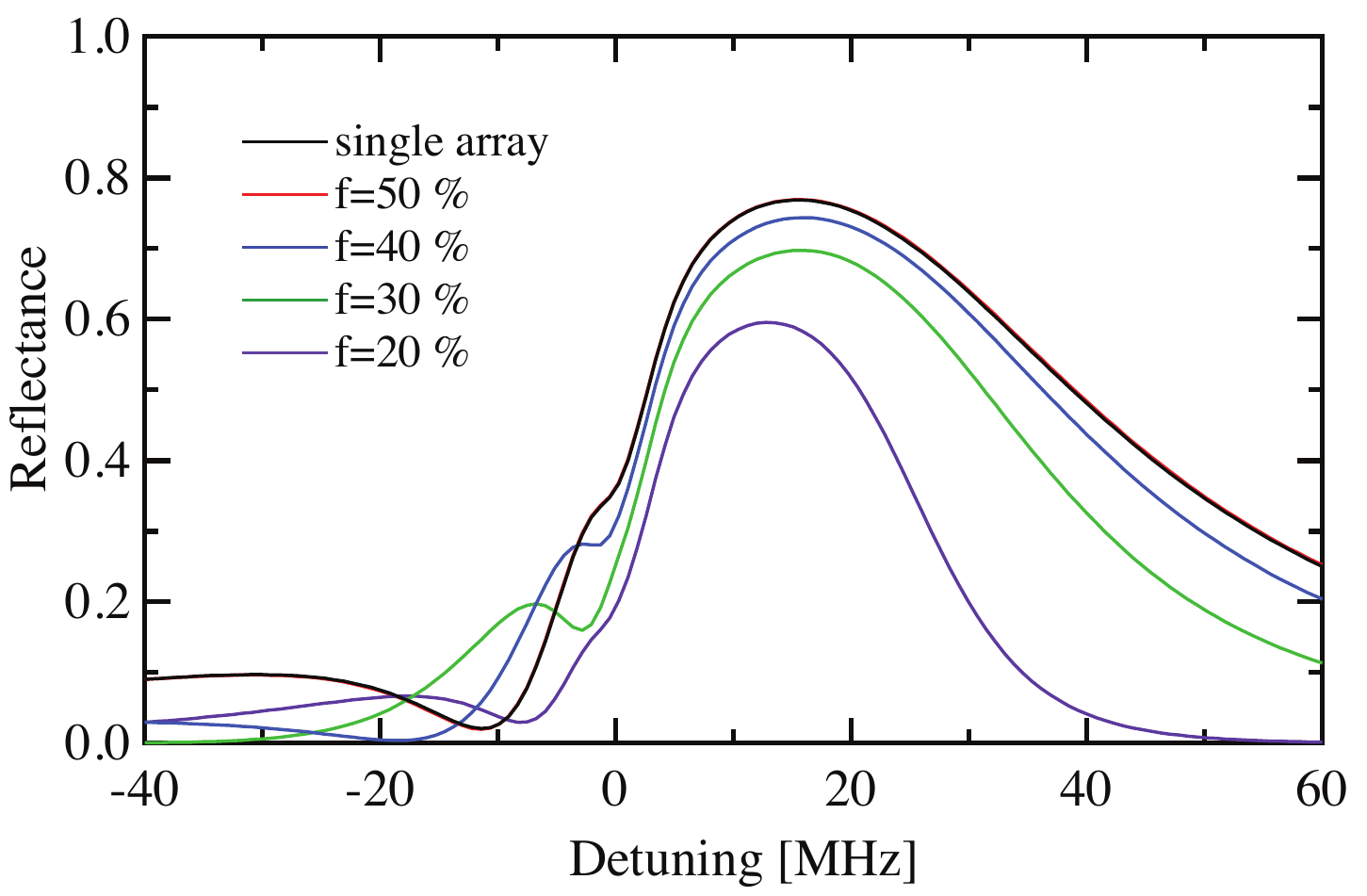}
\caption{Simulated reflection spectra for a probe quasilinearly polarized along the y-direction (symmetric decay rates) and a given filling factor $f$ of each trapping site. We consider two parallel arrays. The trap detuning is $\Delta\lambda=0.2$ nm, the number of atoms $N=2000$ and the decay rate $\Gamma_{\textrm{1D}}/\Gamma_0=0.007$. The spectra are averaged over 15 realizations.}
\label{fig3_SI}
\end{figure}

\section{S3. Effect of disorder induced by the filling factor}

The filling factor of the trapping sites can induce randomness in the distribution of the atoms across the lattice. We consider here two parallel lines of atoms, as obtained with a nanofiber, with at most one atom per site. Given a filling factor per site $f$, at each trapping axial position, there are either no atom at all, one atom on the upper chain, one atom on the lower chain or one atom in both upper and lower chains. This situation can be simulated by a single random array with, for each site, either no atom (probability $(1-f)^2$), one atom (probability $2f(1-f)$) or two atoms (probability $f^2$). As shown on Fig. \ref{fig3_SI}, a filling factor of 50\%, which is the ideal case in the collisional blockade regime, leads here to the same spectrum than a full single array for the same number of atoms. When the filling factor decreases, the spectrum is narrower and the maximum reflection decreases. In the main manuscript, a filling factor $f=0.3$ was used to fit the experimental data.

In the inset of figure 4 of the main manuscript, we finally provided the maximal reflection as a function of the number of remaining trapped atoms.  For this simulation, we started from two parallel arrays with a filling factor $f=0.3$ and subsequently added random loss of the atoms. If one denotes $1-\eta$ the probability for an atom to be lost, the new probability for having one atom is $2\eta f(1-\eta f)$ and for two atoms $\eta^2f^2$.

\begin{figure}[t!]
\includegraphics[width=0.95\columnwidth]{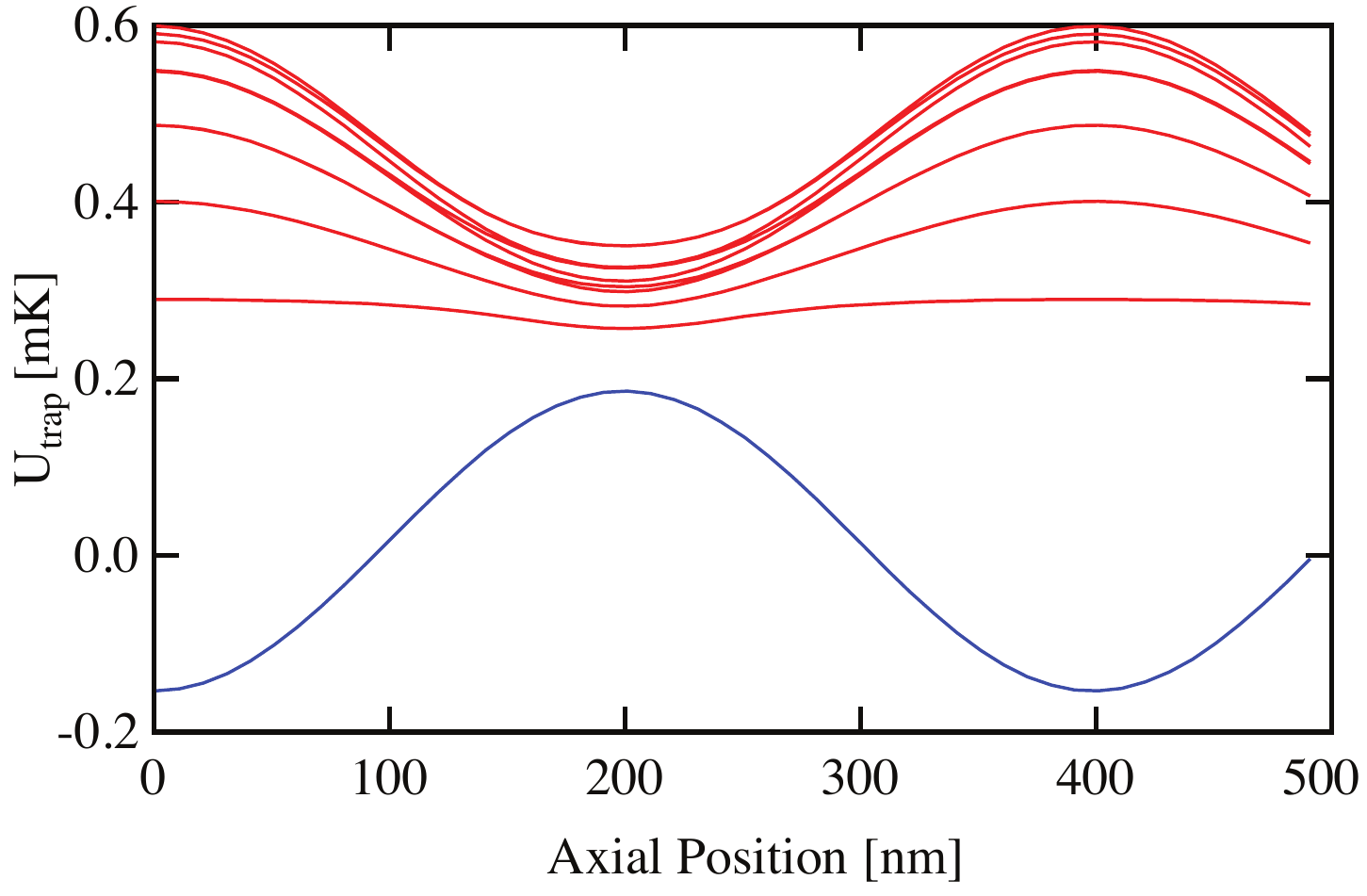}
\caption{Adiabatic trapping potential $U_{trap}(z)$ for the near resonant trap for Cs atoms outside a cylindrical waveguide of 200 nm radius. Here, $\Delta\lambda=0.2$~nm and the red-detuned trapping power is $2\times1.9\mu\rm{W}$. The power of each blue-detuned fields at 686.1 nm and 686.5 nm is 4 mW. 
Blue line corresponds to the axial potential for the ground state manifold $6S_{1/2}~F=4$, while the red lines correspond to the substates of the excited state $6P_{3/2}~F'=5$.}
\label{fig4_SI}
\end{figure}

\section{S4. Trapping potential and effect of disorder induced by imperfect axial localization}

Following the calculation presented in \cite{Lacroutea, LeKienEPJDa}, we calculated the trapping potential and trapping frequency in the axial direction. Figure \ref{fig4_SI}  shows the axial dependence of the trapping potential $U_{\rm{trap}}(x_0,y=0,z)$ for atoms trapped at a distance $x_0=$ 234 nm from the fiber surface. The $U_{\rm{trap}}$ values for the $\{6\rm{S}_{1/2},\rm{F}=4\}$ ground state manifold and $\{6\rm{S}_{1/2},\rm{F}'=5\}$ excited state manifold are plotted as blue and red lines respectively. For this calculation, we use $\Delta\lambda=0.2~\rm{nm}$ and a power of $2\times1.9~\mu\rm{W}$ for the red detuned beams, and a power of $2\times4~\rm{mW}$ for the blue detuned beams. The potential depth at minimum is $U_{\rm{min}}=-0.15$ mK and the trap frequency in the axial direction is $\nu_{z}/2\pi=258~\rm{kHz}$.

In the case of $\Delta\lambda=0.12~\rm{nm}$ and a trapping power of $2\times1~\mu\rm{W}$, the potential depth at minimum is $U_{\rm{min}}=-0.1~\rm{mK}$ and the axial trap frequency is reduced to $\nu_{z}/2\pi=215~\rm{kHz}$. Both axial frequencies for this near resonant trap are comparable with previous nanofiber-based traps \cite{nanofiber2a, Vetscha}.

Using these calculated frequencies, we compute the root mean square of the spatial spread  in the axial direction in the harmonic approximation:
\begin{equation}
\sigma_{z}=\sqrt{(k_{B}T)/(m_{Cs}\nu_z^{2})},
\end{equation}
where $k_{B}$ is Boltzmann constant, $T$ is the temperature, $m_{Cs}$ is the Cesium atomic mass and $\nu_z$ is the axial frequency of the trap. In our case, the temperature of the atomic cloud is estimated to $T=20$ $\mu \rm{K}$ from a time-of-flight measurement. Thus, $\sigma_{z}$ is equal to $22$ $\rm{nm}$ and $26$ $\rm{nm}$ for  $\Delta\lambda=0.2~\rm{nm}$ and $\Delta\lambda=0.12~\rm{nm}$ respectively. For atoms cooled in the ground state, the axial spread would be $12$ $\rm{nm}$ and $13$ $\rm{nm}$.

\begin{figure}[t!]
\includegraphics[width=0.95\columnwidth]{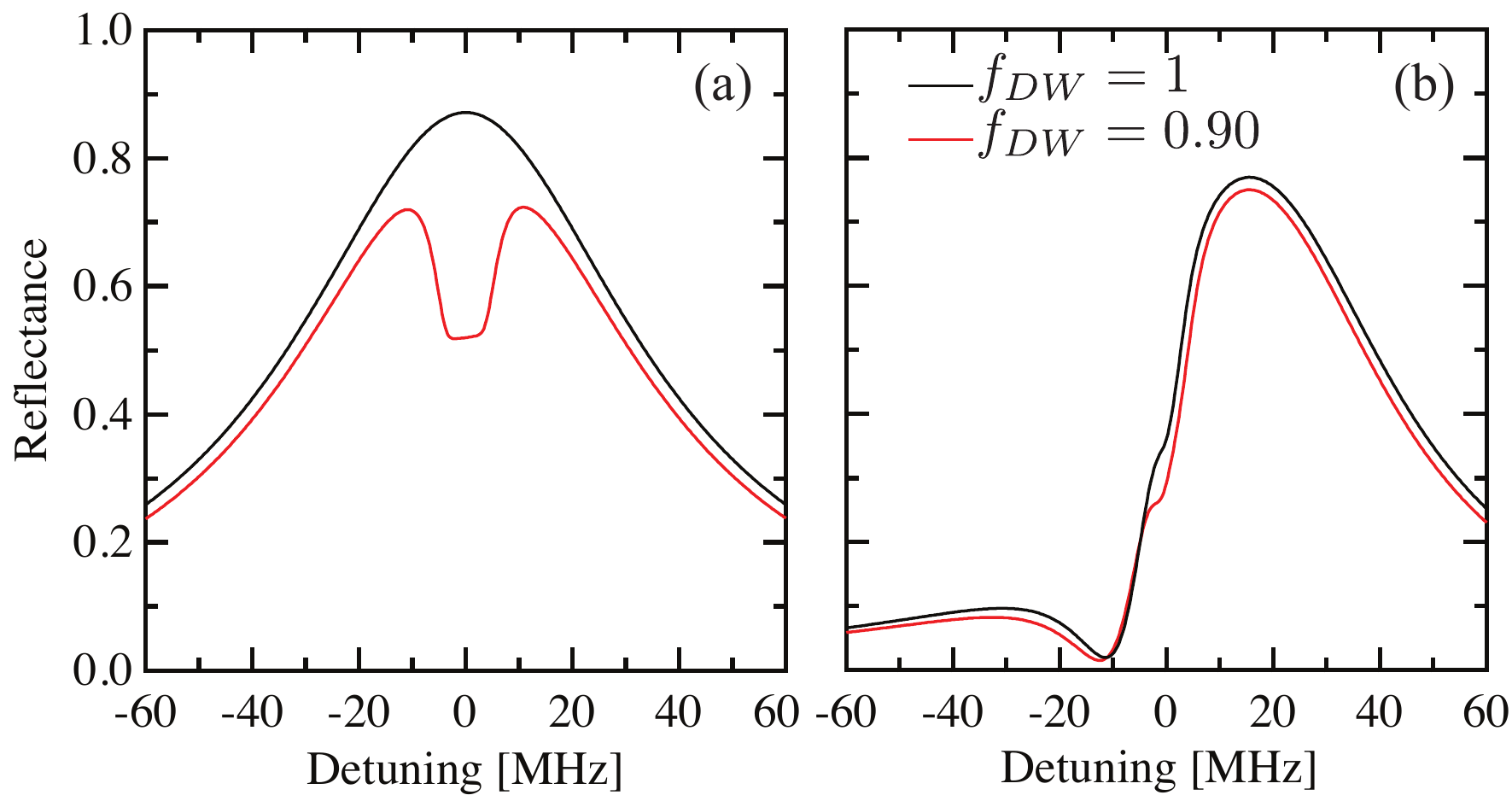}
\caption{Simulated reflection spectra for a probe quasilinearly polarized along the y-direction (symmetric decay rates) with and without spread of the atoms in the potential wells. (a) corresponds to on-resonance trap while (b) corresponds to our experimental case with $\Delta\lambda=0.2$ nm. The Debye-Waller factor of 0.9 corresponds here to $\sigma_z\sim 22$ nm, as estimated from the temperature of the atoms and the simulated trap axial frequency. The number of atoms is $N=2000$ and the decay rate is $\Gamma_{\textrm{1D}}/\Gamma_0=0.007$. The spectra are averaged over 15 realizations.}
\label{fig5_SI}
\end{figure}

The reduction of Bragg reflected intensity due the spread of the atomic position in the potential wells is usually estimated by a so-called Debye-Waller factor \cite{fdwa,Deutscha}. This factor is given by 
$f_{DW}=e^{-4k^2 \sigma_z^2},$ where $k$ is the guided mode wavenumber. In our case, we find values of $0.89$ and $0.85$ for $\Delta\lambda=0.2~\rm{nm}$ and $\Delta\lambda=0.12~\rm{nm}$ respectively  ($0.96$ for atoms cooled to the motional ground state). In Fig. \ref{fig5_SI}, we provide the simulated spectra taking into account a Gaussian spread $\sigma_z$ for the atom position. As it can be seen, this close-to-unity Debye Waller factor has a very limited effect in our configuration where Bragg reflection is obtained out of resonance while it would be an important factor close to resonance \cite{Slama2006a}. We note that the reduction at resonance is larger than the Debye-Waller factor, which is valid in the single-scattering regime.

\end{document}